\documentclass[conference]{IEEEtran}
\IEEEoverridecommandlockouts
\usepackage{amsmath,stfloats}
\usepackage{graphicx}
\include{psfig}
\usepackage{epsfig}

\usepackage{array}
\usepackage{psfrag}
\usepackage{amssymb}
\usepackage{mathdots}
\include{psfig}
\usepackage{color}
\usepackage{pstricks,pst-node,pst-text,pst-3d,pst-plot}
\usepackage{psfrag}
\usepackage{times}
\usepackage{tikz}
\usetikzlibrary{arrows,automata}

\usepackage{epstopdf}
\usepackage{epsf}
\usepackage{epsfig,latexsym,amsmath,epsf,amssymb,amsfonts}

\begin{document}

\title{Improved Spectrum Mobility using Virtual Reservation in Collaborative Cognitive Radio Networks}


\author{\IEEEauthorblockN{Ayman T. Abdel-Hamid, Ahmed H. Zahran\IEEEauthorrefmark{1} \thanks{\IEEEauthorrefmark{1}Ahmed H. Zahran and Tamer ElBatt are also affiliated with  EECE Dept., Faculty of Engineering, Cairo University.}and Tamer ElBatt\IEEEauthorrefmark{1}}
\IEEEauthorblockA{Wireless Intelligent Networks Center (WINC) \\
Nile University, Cairo, Egypt.\\ Email: {\tt
ayman.tharwat@nileu.edu.eg, ahzahran@ieee.org, telbatt@ieee.org}}
}

\maketitle
\footnotetext[1]{This work was funded by the Egyptian National Telecommunication
Regulatory Authority (NTRA).}
\vspace{-1mm}

\begin{abstract}
Cognitive radio technology would enable a set of secondary users (SU) to opportunistically use the spectrum licensed to a primary user (PU). On the appearance of this PU on a specific frequency band, any SU occupying this band should free it for PUs. Typically, SUs may collaborate to reduce the impact of cognitive users on the primary network and to improve the performance of the SUs. In this paper, we propose and analyze the
performance of virtual reservation in collaborative cognitive
networks. Virtual reservation is a novel link maintenance
strategy that aims to maximize the throughput of the cognitive
network through full spectrum utilization. Our performance evaluation shows significant improvements not only in the SUs blocking and forced termination probabilities but also in the throughput of cognitive users.\\

\end{abstract}
\IEEEpeerreviewmaketitle

\section{Introduction} \label{intro}
Cognitive Radio (CR) \cite{akyildiz2006next} is proposed to improve the utilization of the spectrum that is shown to be underutilized in several measurement studies. Several portions of the scarce spectrum are assigned to a set of primary users (PUs) that are not efficiently using this valuable resource. CR enables another set of users, commonly known as secondary users (SUs), to opportunistically transmit their data using the same spectrum provided that such transmission should have no or very limited impact on the PU communications.\\
In order to minimize the impact of SUs on PUs, SUs typically senses the medium before using it. Spectrum sensing is generally categorized into reactive and proactive sensing \cite{ghasemi2008spectrum}. In the former, sensing is performed on-demand basis, when a SU has data to transmit, while in the latter, SUs periodically sense a portion of the spectrum even if they have no data to send. Note that the proactive approach aims to minimize the sensing delay and is typically favoured by delay-sensitive applications (e.g., Real Time Applications). However, reactive sensing is more favourable when the energy efﬁciency of SUs is of concern. Sensing can also be categorized as collaborative and non-collaborative sensing \cite{ben2009cooperative}. In the former, a set of SUs cooperates to sense a band of interest and share their sensing outcome and collaborate for the benefit of each other, while in non-collaborative schemes, SUs individually sense the medium and their transmissions may affect coexisting SUs. In cooperative sensing, the collaboration is typically performed using a common control channel over which control information is shared. In our model, we are mostly concerned with real-time applications which need to guarantee minimum bandwidth.\\
Generally speaking, the sensing process is performed via one of three approaches: Interrupted sending, Dynamic Frequency Hopping (DFH) and partial sensing \cite{willkomm2010efficient}. Interrupted sending means that secondary data
transmission has to be periodically interrupted to perform
sensing. On another side, SUs may follows the idea of performing sensing and data
transmission in parallel in different PU bands and switching
them cyclicly (DFH). Typically, this needs a complex radio front end. The alternative way is Partial Sensing, where not the whole PU band is scanned for constructing one SU link, but some part is always left idle to perform sensing in parallel by other cooperative inactive SUs. Intuitively, perfect sensing is worth investigating; as SU need to efficiently discover depleted regions as well as avoiding complexity and spectral overhead \cite{kim2008efficient}.\\
In CR systems, the sudden appearance of  a PU on a band occupied by a SU triggers the cognitive user to leave this band. The SU would then try to reacquire the medium through one of the following three actions: (1) staying in the original
channel and postpones its transmission until the PU finishes, (2) selecting a channel from a list of previously
sensed channels to replace his purloined channel (predetermined spectrum handoff) or (3) switching to a certain channel after immediate sensing (sensing-based spectrum handoff). If the SU fails to reacquire the spectrum, it is forced to terminate its session \cite{akyildiz2008survey}. In \cite{wang2008performance} the effective data rate for the sensing-based spectrum handoff is higher than the pre-determined channel 
list spectrum handoff in the case of the high primary user’s traffic load.\\
In order to reduce the event of forced termination, several link maintenance strategies \cite{willkomm2005reliable} are proposed in the literature to ensure a minimum bandwidth is available for SU with a very high probability, even in case of spectrum handoff. Two main link reconfiguration approaches \cite{willkomm2010efficient} are observed in the literature including over-provisioning and resource reservation. Over provisioning adds redundancy to SU transmission such that if a portion of its occupied spectrum is compromised, the communication over the remaining occupied portion would still be sufficient to successfully communicate the message. In resource reservation schemes, a portion of the spectrum is kept unused as a back up for spectrum handoffs. One way to practically manage resource reservation schemes is to perform admission control on SU traffic.\\
Optimal admission was first introduced in cellular networks to set optimal policies for 
accepting new or handoff calls to minimize the blocking possibilities of both types of calls. That was introduced in \cite{ramjee1997optimal} through (Guard channel) and (Fractional Guard channel) policies. Moreover, Kumar and Tripathi in \cite{kumar2009adaptation} proposed a new preemptive handoff scheme such that a degree of protection is
offered to the ongoing calls in the handoff region. Yet, the cellular handoff model is completely different from the $CRN_{s}$ one; in $CRN_{s}$ case the existence of the highly prioritized PU gives him the authority to lord it over $SU_{s}$. In \cite{pacheco2009optimal}, an optimal admission policy is proposed to balance the trade-off that arises between blocking new sessions and handover requests in $CRN_{s}$ context.\\
In this paper, we introduce the concept of virtual reservation in collaborative CR networks. In this context, virtually reserved channels are channels reserved for handoff requests but active SUs are allowed to use them, i.e. the channels are not left unused until a spectrum handoff is imminent. On the appearance of a PU, the active SU(s) collaborate to ensure that a minimum amount of resources are provided to each user. Adopting virtual reservation implies full spectrum utilization (FSU) as it is the case in over-provisioning link maintenance schemes. We develop a performance evaluation framework for the proposed virtual reservation scheme to estimate different key performance indcies (KPIs) such as forced termination and blocking probabilities and the system throughput. Additionally, we develop a framework to optimally determine the number of virtually reserved channels that compromise an existing tradeoff between different KPIs. Our results show a significant improvement of the derived KPIs in comparison to non-collaborative CR networks as well as existing collaborative approach \cite{lee2008optimal} that does not adopt FSU.\\
The rest of this paper is organized as follows. In Section \ref{modeldesc}, the primary and secondary user's models are defined. Moreover, our model employing FSU and Virtual Reservation is introduced in this section. Then, we analyse our main contributions in Section \ref{bvsft} as a result of employing FSU and Virtual Reservation. After that, in Section \ref{num} the numerical results including the formulation and solution of an unconstrained optimization problem, in order to derive the optimal reservation values, are interpreted. Eventually, we draw some related conclusions in Section \ref{conclusion}.

\section{System Model} \label{modeldesc}
In this section, we first present the system description followed by an overview for virtual reservation scheme. We then present the system Markovian model followed by a framework for estimating the main performance metrics. 
\subsection{System Description}
We consider a system with C channels, divided into M blocks each
having N channels. PUs arrive according to a Poisson process with an average arrival rate $\lambda_{p}$. Each  PU  session takes over a  band  of  N  channels for an exponential period with rate $\mu_{p}$. SU traffic arrival is Poisson with rate $\lambda_{s}$. SUs are assumed to have a minimum quality of service (QoS) requirements represented by a minimum number of traffic channels, denoted as $C_{min}$. These $C_{min}$ channels correspond to an exponential service time with rate $\mu_{s}$. Additionally, SU traffic is assumed to have a greedy nature, i.e. SU traffic can use more channels by which SUs can reduce their service time. In this case, the service rate is scaled with a factor corresponding to the ratio of the allocated channels to $C_{min}$.\\
SUs adopt collaborative sensing and resource allocation schemes. Each SU is assigned a set of channels to scan, such that at the end of a sensing period, the system has a full information about the free channels in the band of interest shared through a common control channel \cite{lee2008optimal}. The resource allocation scheme equally distributes idle channels among active SUs such that each SU is guaranteed at least its minimum bandwidth requirements, i.e. $C_{min}$. Note that extra available channels are equally shared among active SUs using schemes such as time sharing for a fair resource distribution.
\subsection{Virtual Reservation Framework}
In the proposed virtual reservation scheme, a number of channels, denoted as $r$, are set-aside for SUs spectrum hand-offs and new SUs are not allowed access to those channels. It is worth noting that these channels are not kept free, as in traditional link maintenance schemes \cite{zhu2007analysis}, but they are utilized by active SUs to enhance the throughput of cognitive users.
Obviously, all idle channels are occupied by SUs, hence the spectrum of interest would be fully occupied once there exist at least one active SU.\\
In our model, we assume that the cognitive network admits a new user if the number of idle channels less the number of reserved channels are sufficient to accommodate the minimum requirements of all active SUs in addition to the new user requirement. This admission condition can be expressed as
\begin{equation}
C-Nn_{p}-r-C_{min}(n_{s}+1)\geq 0,
\end{equation}
where $n_{p}$ and $n_{s}$ represent the number of active primary and secondary users, respectively. This condition guarantees a minimum QoS requirement for collaborating SUs. Note that if the new SU admission condition is not satisfied, the new SU is blocked. Additionally, on the appearance of a PU, one or more SUs may be forced to terminate if the remaining idle channels, after the PU activity, do not suffice the minimum QoS requirements of active SUs. Hence, the forced termination event occurs when
\begin{equation}
C-N(n_{p}+1) < C_{min}n_{s}.
\end{equation}
In the case of forced termination, the number of active SUs drops to
\begin{equation}
\grave{n_{s}}=\lfloor\frac{C-N(n_{p}+1)}{C_{min}}\rfloor
\end{equation}
Hence, after the admission of a new PU, the number of active SUs can be expressed as
\begin{equation}
n_{s}=
\begin{cases}
n_{s}, &\text{if $C-N(n_{p}+1)\geq C_{min}n_{s}$;}\\
\grave{n_{s}}, &\text{if $C-N(n_{p}+1)< C_{min}n_{s}$.}     \label{relation1}
\end{cases}
\end{equation}
We assume that the users that are forced to terminate are randomly chosen from the pool of active SUs.

\subsection{The System Markovian Model}
\begin{figure}[hbtp]
\centering
\begin{tikzpicture}[>=stealth',shorten >=1 pt,auto,node distance=2.5 cm,semithick]
\tikzstyle{every state}=[text width=1 cm ,minimum size=35 pt,fill=white,draw=black,thick,text=black,scale=0.8]

\node[state]         (I1)  {$n_{p},n_{s}$};
\node[state,dashed]         (J1) [right of=I1,xshift=0.7cm] {$n_{p}+1,n_{s}$};
\node[state]         (L2) [above of=I1] {$n_{p},n_{s}+1$};

\node[state]         (B1) [below of=I1] {$n_{p},n_{s}-1$};

\node[state]         (E1) [left of=I1,xshift=-0.7cm] {$n_{p}-1,n_{s}$};

\node[state,dashed]         (Q) [below of=I1,xshift=2.5cm,yshift=-0.2cm] {$n_{p}+1,\grave{n_{s}}$};

\begin{scope}[every node/.style={scale=0.8}]

\path (I1) edge  [bend left,->] node[below] {$n_{p}\mu_{p}$} (E1);
\path (E1) edge  [bend left,->] node[below] {$\lambda_{p}$} (I1);

\path (J1) edge  [bend left,->] node[above] {$(n_{p}+1)\mu_{p}$} (I1);
\path (I1) edge  [bend left,->] node[below] {$\lambda_{p}$} (J1);

\path (I1) edge  [bend left,->] node[left] {$n_{s}\mu_{s}$} (B1);
\path (B1) edge  [bend left,->] node[left] {$\lambda_{s}$} (I1);

\path (L2) edge  [bend left,->] node[right] {$(n_{s}+1)\mu_{s}$} (I1);
\path [dashed] (I1) edge  [bend left,->] node[left] {$\lambda_{s}$} (L2);

%\path (Q) edge  [bend left,->] node[right] {$(n_{s}+1)\mu_{s}$} (I1);
\path (I1) edge  [->] node[right] {$\lambda_{p}$} (Q);

\end{scope}
\end{tikzpicture}
\caption{State Transition Rates for the 2D Markov chain}
\label{fig1}
\end{figure}
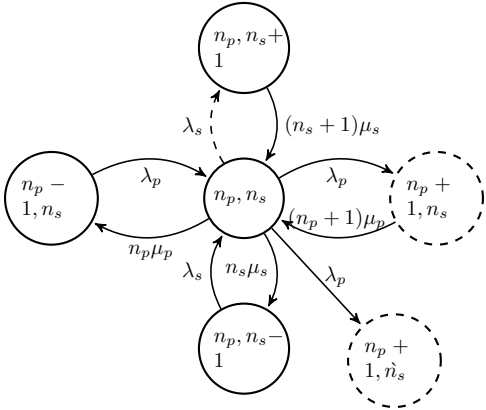

In this section, we analyze the performance of virtual reservation for greedy applications in collaborative cognitive networks. The analysis is based on a multidimentional Markov chain whose states are described by an integer pair ($n_{p},n_{s}$).
% % % % % % % % % % % % % % % % % % % % % % % % % % % % % % % % % % % % % % % % % % % % % % % % % % % % % % % % % %
Figure \ref{fig1} shows the different states and transition rates of the continuous-time Markov chain (CTMC) that models the system dynamics.

% % % % % % % % % % % % % % % % % % % % % % % % % % % % % % % % % % % % % % % % % % %
% % % % % % % % % % % % % % % % % % % % % % % % % % % % % % % % % % % % % % % % % % % %
As shown in Figure \ref{fig1}, the arrival and departure of a PU are captured by horizontal transitions except for a PU arrival causing forced termination event (diagonal transition). The maximum number of PUs that can coexist with $n_{s}$ users in the system, denoted as $U_{pn_{s}}$,  can be expressed as

\begin{equation}
\begin{aligned}[c]
{U}_{pn_{s}}&= \lfloor \frac {NM-n_{s}C_{min}}{N}\rfloor \\
&= \lfloor M-\frac{n_{s}}{N}C_{min}\rfloor           \;\;\;\;\;\;\;\;   n_{s}=1,2,... \label{relation2}
\end{aligned}
\end{equation}

In case that the number of PUs ($n_{p}$) exceeds ${U}_{pn_{s}}$, the chain is transited from state ($n_{p},n_{s}$) to  the dashed state ($n_{p}+1,\grave{n_{s}}$), i.e. a number of SUs are dropped of the system ($U_{d}(n_{p},n_{s})$). Note that, $U_{d}(n_{p},n_{s})$ is not necessarily one user, however it's upperbounded by $\lceil\frac{N}{C_{min}}\rceil$. In contrast to traditional reservation models, in our model $U_{d}$ is a deterministic number known for each state ($n_{p},n_{s}$); that is the FSU implies the full free spectrum occupation by SUs. Thus, whenever some resources are lost due to a PU appearance, the whole band of interest is scanned cooperatively and reported for efficient resource management.
Similarly, the arrival and departure of SUs are depicted using vertical transitions, except for forced termination conditions. Typically, if there exist $n_{p}$ PUs in the system, the maximum number of coexisting SUs, denoted as $U_{sn_{p}}$, can be expressed as
\begin{equation}
{U}_{sn_{p}} = \lfloor \frac {NM-(Nn_{p}+r)}{C_{min}}\rfloor     \;\;\;\;\;\;\;\;   n_{p}=1,2,...M  \label{relation3}
\end{equation}
Any SU arrival beyond this limit is not allowed to share the system resources; i.e. blocked. Thus, in these blocking states, the upward transition ($\lambda_{s}$) is not feasible.
However, the chain dynamics and the resource management rules create states in which more SUs may coexist in the system depending on the order of SU and PU arrival. These states occur when the number of SUs approaches the limit in (\ref{relation3}) and an additional PU arrives. In this case, the system would try to reduce the number of forced terminated SUs while temporarily not guaranteeing the number of virtually reserved channels.
To further illustrate these cases, consider a sample system $3\times 4$ with $C_{min}=2$ and $r=2$. If the system  is in state (1,$n_{s}$), the maximum possible number of SUs is $n_{s}=3$. However, system can reach the state (1,4) when the system is in (0,4) and then encounters a PU arrival ($\lambda_{p}$ transition in the horizontal dimension).\\
Based on this analysis, the transition rate, denoted as $q_{h\tilde{h}}$, from state $h(n_{p},n_{s})$ to another state $\tilde{h}(\tilde{n}_{p},\tilde{n}_{s})$ can be expressed as

\begin{equation}
\begin{aligned}[c]
q_{h\tilde{h}}=
\begin{cases}
\lambda_{p} ,  &\text{$\tilde{n}_{s}=n_{s}, \tilde{n}_{p}=n_{p}+1$ and}\\
&\text{$N\tilde{n}_{p}+C_{min}\tilde{n}_{s}\leq C$;}\\
\lambda_{p},   &\text{$\tilde{n}_{s}=\grave{n_{s}}, \tilde{n}_{p}=n_{p}+1$ and}\\ 
&\text{$N\tilde{n}_{p}+C_{min}\tilde{n}_{s}> C$;}\\
\lambda_{s},   &\text{$\tilde{n}_{s}=n_{s}+1$, $\tilde{n}_{p}=n_{p}$ and}\\ 
&\text{$Nn_{p}+C_{min}\tilde{n}_{s}+r\leq C$;}\\
n_{p}\mu_{p},       &\text{$\tilde{n}_{p}=n_{p}-1$ and $\tilde{n}_{s}=n_{s}$;}\\
\frac{N(M-n_{p})}{C_{min}}\mu_{s},       &\text{$\tilde{n}_{s}=n_{s}-1$ and $\tilde{n}_{p}=n_{p}$;}\\
0,                       &\text{Otherwise}   \label{relation6}
\end{cases}
\end{aligned}
\end{equation}

The stationary distribution of this chain, denoted as $\pi(h)$, can be estimated using the global balance equation and the normalization equation.
\subsection{Performance Evaluation}
The first performance metric of interest is the blocking probability for SUs requests, denoted as ($P_{B}^{s}$), is estimated as the sum of the stationary probabilities of SU blocking states and can be expressed as
\begin{equation}
P_{B}^{s} =\sum_{n_{p}=0}^M\sum_{n_{s}=0}^{U_{sn_{p}}} I(Nn_{p}+C_{min}(n_{s}+1)+r,NM)\pi(n_{p},n_{s})   \label{relation11}
\end{equation}
where I(x,y) represents an indicator function and is expressed as
\begin{equation}
I(x,y)= 
\begin{cases}
1, &\text{if $x\geq y$;}\\
0, &\text{if $x< y$}      \label{relation10}
\end{cases}
\end{equation}

Also the forced termination probability of active SUs ($P_{FT}^{s}$) is estimated as the ratio of SU forced termination rate to the rate of SU admission, \cite{ahmed2009comments}, and can be expressed as

\begin{equation}
P_{FT}^{s} =\frac{\sum_{n_{p}=0}^M\sum_{n_{s}=0}^{U_{sn_{p}}} {U}_{d}(n_{p},n_{s})q_{n_{p}+1,\grave{n_{s}}}^{n_{p},n_{s}}\pi(n_{p},n_{s})}{(1-P_{B}^{s})\lambda
_{s}}    \label{relation12}
\end{equation}
where $q_{n_{p}+1,\grave{n_{s}}}^{n_{p},n_{s}}$ is the transition rate from state ($n_{p},n_{s}$) to state ($n_{p}+1,\grave{n}_{s}$). In our model, this transition rate is equal to $\lambda_{p}$, reflecting the event of a PU appearance.\\
Last but not least, we can define the SU throughput as the average number of channels which are usefully used by a SU in the system $C_{avg}$. $C_{avg}$ is estimated, using the Markov Reward Model (MRM) \cite{bolch2006queueing}, by defining the number of channels assigned per SU as a reward rate for each state $h(x,y)\in \grave{S}$, which represents the set of all states with non-zero SU occupancy. Hence, $C_{avg}$ can be estimated as

\begin{equation}
C_{avg} =\sum_{n_{p}=0}^M\sum_{n_{s}=1}^{U_{sn_{p}}} (\frac{N(M-n_{p})}{n_{s}}) \pi(n_{p},n_{s})        \label{relation13}
\end{equation}
% % % % % % % % % % % % % % % % % % % % % % % % % % % % % % % % % % % % % % % % % % % % % % % % % % % %
\section{Analysis of FSU and Virtual Reservation } \label{bvsft}
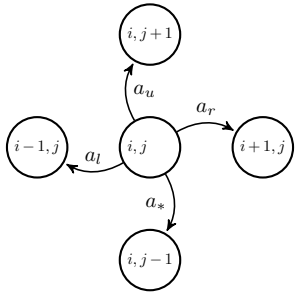
\begin{figure}[hbtp]
\centering
\begin{tikzpicture}[>=stealth',shorten >=1 pt,auto,node distance=2.5 cm,semithick]
\tikzstyle{every state}=[text width=1 cm ,minimum size=35 pt,fill=white,draw=black,thick,text=black,scale=0.6]

\node[state]         (AA)        {$i,j$};
\node[state]         (AB) [right of=AA] {$i+1,j$};
\node[state]         (AC) [left of=AA] {$i-1,j$};
\node[state]         (AD) [above of=AA] {$i,j+1$};
\node[state]         (AE) [below of=AA] {$i,j-1$};

\begin{scope}[every node/.style={scale=0.8}]

\path (AA) edge  [bend left,->] node[above] {$a_{r}$} (AB);
\path (AA) edge  [bend left,->] node[above] {$a_{l}$} (AC);
\path (AA) edge  [bend left,->] node[right] {$a_{u}$} (AD);
\path (AA) edge  [bend left,->] node[left] {$a_{*}$} (AE);

\end{scope}
\end{tikzpicture}
\caption{A general sample state to apply Drift Analysis}
\label{fig2}
\end{figure}

In this section, we rigorously prove that the adoption of FSU and Virtual Reservation approaches enhances link maintenance in terms of both the blocking and forced termination probabilities ($P_{B}^{s},P_{FT}^{s}$) for SUs.\\
We use the notion of drift as a measure of stability of the Markov chain \cite[Ch.4]{bertsekas1992data}. For discrete-time Markov chains, the drift is the expected change in state given that the chain is in a certain state $i$, i.e. for one-dimensional Markov it is expressed as
\begin{equation}
D_{i} = E\{X_{n+1}-X_{n}|X_{n}=i\} = \sum\limits_{K=-i}^{\infty} K P_{i,i+K}        \label{relation133}
\end{equation}
where $P_{i,i+k}$ is the single-step transition probability derived from the Embedded Markov chain (EMC) of the corresponding CTMC constructed in Section \ref{modeldesc}. We define $\hat{S}$ as the set of states connected to state $h=(i,j)$ as shown in Figure \ref{fig2}. Hence, the transition probabilities of the EMC are expressed as, $\mathbf{p_{h\tilde{h}}=\frac{q_{h\tilde{h}}}{\sum\limits_{\forall \tilde{h}\in \hat{S}}q_{h\tilde{h}}}}$, \cite[Ch.6]{ross2009introduction}, where $p_{h\tilde{h}}$ and $q_{h\tilde{h}}$ represent the transition probabilities and rates for the EMC and CTMC, respectively, from state $h$ to $\tilde{h}$.

In Figure \ref{fig2}, a general state is used to examine the stability of the system Markovian model. Assuming that $D_{i,j}$ is the drift for the state $(i,j)$, where negative drift indicates transitions downward or left (stable behavior) and positive drift (unstable behavior) indicates upward or right transitions. Thus, applying the drift analysis by summing over all transitions probabilities directed out of the discrete-time state (i,j), the state drift can be expressed as
\begin{equation}
D_{i,j} = \frac{a_{r}+a_{u}-a_{l}-a_{*}}{\sum\limits_{\forall v\in\{r,u,l,*\}}a_{v}}        \label{relation134}
\end{equation}
Our baseline model to which we compare our performance is assumed to be similar to the model adopted in \cite{lee2008optimal} in which cognitive users sense the medium in a cooperative manner. However, unlike our model, all spectrum information is delivered to a fusion centre, which allocates only the minimum bandwidth requests to each SU.
Our major difference glows at the downward transition; assigning all free channels to SUs generally leads to a higher utilization per cognitive user than the reference model. Intuitively, at moderate and low traffic load, $a_{*}$ steps up, driving $D_{i,j}$ to be more negative than \cite{lee2008optimal}. More negativity indicates more stability, that is, the system tends usually to stay faraway from the blocking and forced termination states, i.e. border states. Hence, we argue that our model including the FSU approach combined with Virtual Reservation introduces an efficient mechanism for a reliable link maintenance. It's worth noting that, even if we compare to a non-collaborative model, i.e. once a SU detects his minimum requirements of idle channels, he immediately starts his session, the previously mentioned argument for downward transitions remains valid. Also, for all other types of states do not including downward transitions, stability for these states is similar to the traditional model. However, emphasizing on the overall stability of the model, our model outperforms the reference one.
% % % % % % % % % % % % % % % % % % % % % % % % % % % % % % % % % % % % % % % % % % % % % %
\section{Numerical Results and Discussion} \label{num}
In this section, we present numerical results that show the benefits of using virtual reservation with cooperative sensing on the derived performance indicies considering different operating scenarios. We also show how proper configuration for the number of reserved channels (r) contributes to the Blocking-Forced termination optimization. Later on, we analyze the effect of the rate by which PUs are served on system performance. Eventually, we specify how the variation of $C_{min}$ affects both $P_{FT}^{s}$ and $P_{B}^{s}$. The presented results are obtained using Matlab with the default parameters shown in Table \ref{table1} unless otherwise is indicated.

\begin{table}[ht]
\caption{Simulation Parameters} % title of Table
\centering  % used for centering table
\begin{tabular}{|c|c|} % centered columns (4 columns)
\hline                        %inserts double horizontal lines 
%heading
\hline                  % inserts single horizontal line
PU arrival rate($\lambda_{p}$) & 1.3 \\
PU service rate per channel($\mu_{1}=\frac{\mu_{p}}{N}$) & 1   \\
SU service rate per channel($\mu_{2}$) & 0.75  \\
SU loading factor($\rho_{s}=\frac{\lambda_{s}}{\mu_{2}}$) & 0.6 \\ % inserting body of the table
M & 4 bands \\
N & 5 channels/band \\
$C_{min}$ & 2 \\  [1ex]      % [1ex] adds vertical space

\hline %inserts single line
\end{tabular}
\label{table1} % is used to refer this table in the text
\end{table}

\subsection{Throughput, $P_{FT}^{s}$ and $P_{B}^{s}$ for our proposed model}
\begin{figure}[hbtp]
\centering
\includegraphics[width=1\linewidth,height=0.25\textheight]{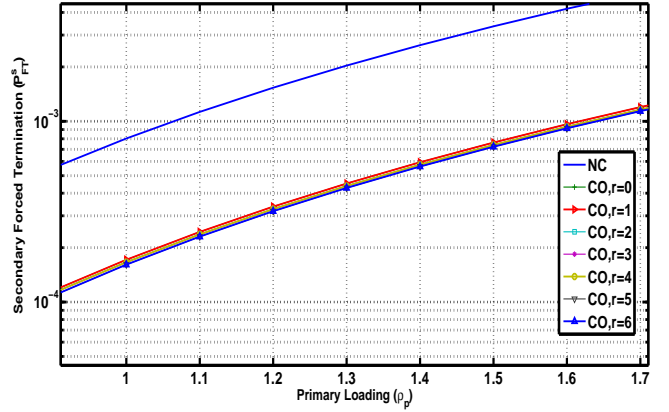}
\caption{Virtual Model-$P_{FT}^{s}$ Vs. primary loading $\rho_{p}$ at various reservation cases}   \label{figcompare1}
\label{fig.4}
\end{figure}

\begin{figure}[hbtp]
\centering
\includegraphics[width=1\linewidth,height=0.25\textheight]{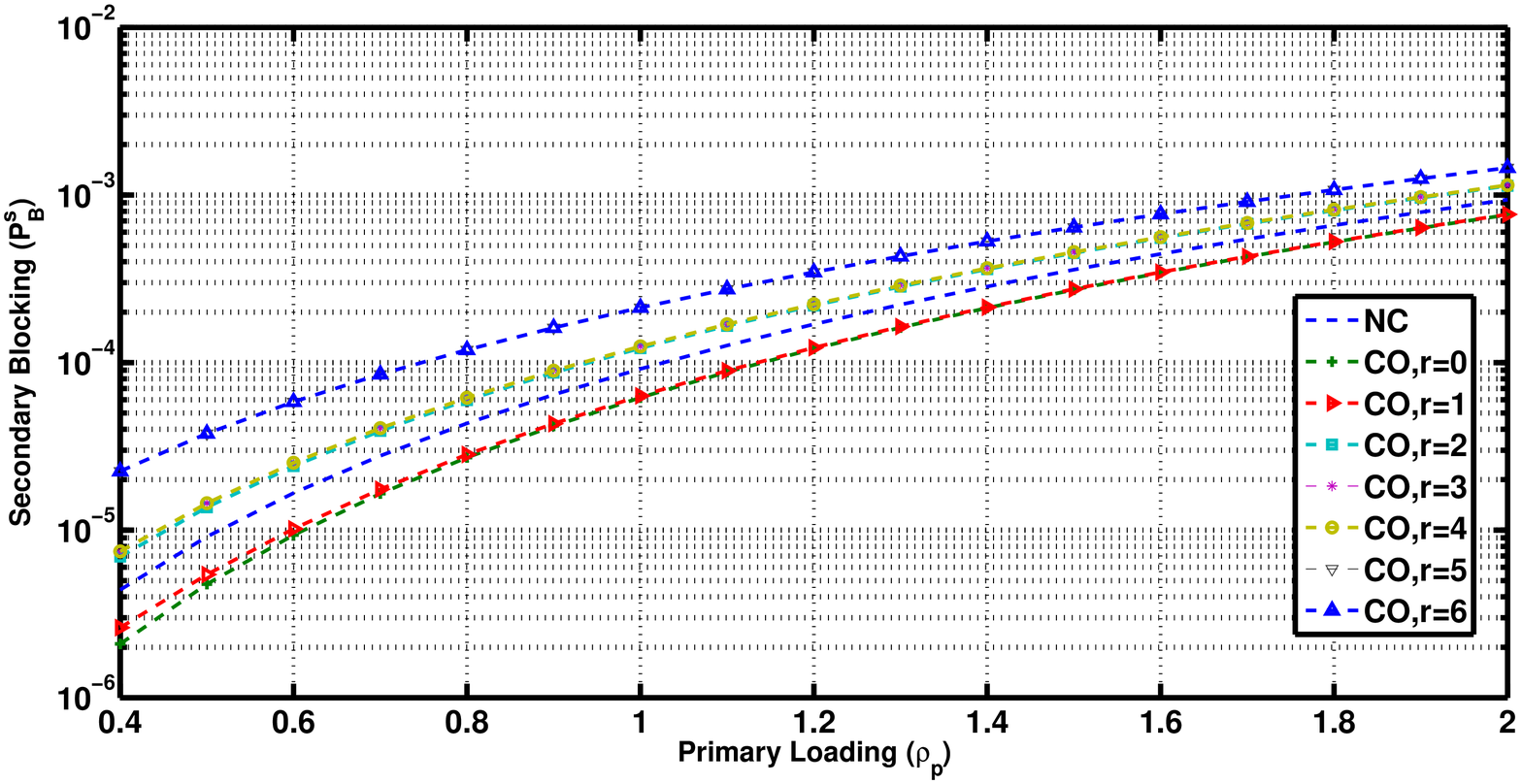}
\caption{Virtual Model-$P_{B}^{s}$ Vs. primary loading $\rho_{p}$ at various reservation cases}   \label{figcompare2}
\label{fig.5}
\end{figure}

\begin{figure}[hbtp]
\centering
\includegraphics[width=1\linewidth,height=0.25\textheight]{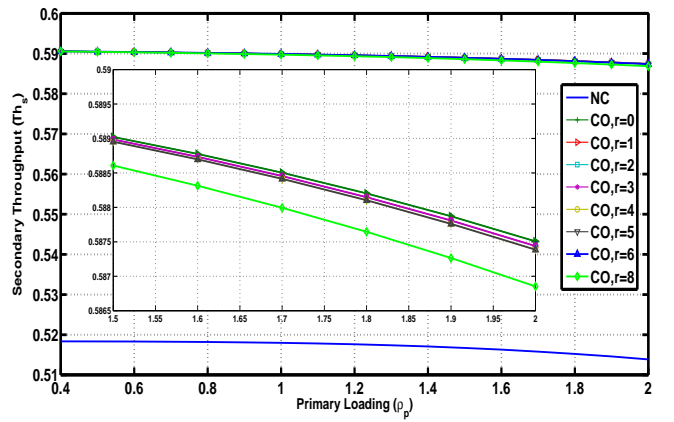}
\caption{Virtual Model-$Th^{s}$ Vs. primary loading $\rho_{p}$ at various reservation cases including an individual plot of the virtual model SU throughput to show the effect of reservation}   \label{figthr}
\label{fig.6}
\end{figure}

In this subsection, we compare the presented scheme to non-cooperative systems in which the user would individually sense the medium and start transmission once $C_{min}$ vacant channels. Note that the sensing process may take longer time in non-cooperative in comparison to cooperative systems. However, such delay analysis is considered a future work. Figure \ref{figcompare1} plots the SU forced termination probability versus the PU loading for different number of virtually reserved channels in cooperative (CO) in addition to non-cooperative (NC) systems. The figure shows a significant reduction in the forced termination probability in comparison to non-cooperative system. A major portion of this gain is due to FSU by comparing NC system with CO system with zero reservation. Further reductions are attained by increasing the number of reservation channels. This additional gain is natural as fewer users would have to terminate their session on the appearance of a PU. Figure \ref{figcompare2} plots the SU blocking probability versus the PU loading for different number of virtually reserved channels in cooperative (CO) in addition to non-cooperative (NC) systems. This figure shows that FSU significantly reduces the blocking probability in comparison to NC systems. This apparent reduction is due to FSU, i.e. in moderate traffic loading, the SU gains benefit of more channels than usual. Consequently, the SU service time is clearly reduced and the system hardly gets congested. However, this reduction naturally decreases as the number of reserved channels increases due the tradeoff between forced termination and blocking probabilities.\\
Figure \ref{figthr} plots the SU throughput versus PU loading for different number of virtually reserved channels in cooperative (CO) in addition to non-cooperative (NC) systems. Clearly, the figure shows an exponential drop in the SU throughput as the PU loading increases. The figure also shows an improvement of 14$\%$ in the SU throughput at unity primary loading. These observations indicate that increasing channel reservation reduces the number of admitted users.  The percentage of time over which the system has no SUs increases after these SUs leave the system. This behavior is magnified when the available resources for SU become limited at high PU loading.

\subsection{The Optimal Virtual Reservation Policy}\label{mincost}
\begin{figure}[hbtp]
\centering
\includegraphics[width=1\linewidth,height=0.25\textheight]{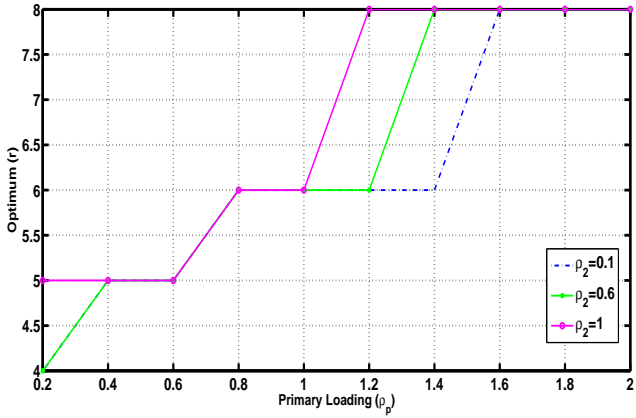} \vspace{-2mm}
\caption{Optimum r as a function of the arrival rate of primary users}  \label{figoptimal}
\label{fig.3}
\end{figure}

From the figures showing cognitive forced termination, blocking probabilities and throughput above, intuitively, at each PU loading condition, there exists an optimal number of reserved channels. Optimality here implies decreasing both SU forced termination and blocking which. Hence, we aim at finding the optimal number of reserved channels via optimizing the forced termination-blocking trade-off, i.e. we can formulate the following problem:
\begin{equation}
\begin{aligned}
& \underset{r}{\text{minimize}}
& & \zeta = \alpha P^{s}_{ft}+P^{s}_{b} \\
& \text{subject to}
& & C-Nn_{p}-C_{min}n_{s}\geq 0          \label{relation15}
\end{aligned}
\end{equation}
where $\alpha\in[0,\infty)$, is a design parameter that controls the weight of $P^{s}_{ft}$ relative to $P^{s}_{b}$ i.e. a
forced termination is $\alpha$ times more costly than the blocking event.\\
Accordingly, solving the previous optimization problem gives back the optimized number of reserved channels at specified PU arrival rate, as we derived in Figure \ref{figoptimal}. The plot includes this relationship at different values of SU utilization($\rho_{s}$). It's obvious that as long as the system gets more congested by PUs ($\lambda_{p}$ increases), active SUs become more threatened to be evicted. Thus, the reservation value must accommodate that change and increase for SUs survival. Note also that at higher SU utilization values ($\rho_{s}$), reservation values increase to keep saving excess users as long as possible. Consequently, choosing the optimal $r$ based on the system dynamics and involved trade-offs drives the system to operate at its optimal state.

\subsection{The effect of PU activity on $P_{FT}^{s}$ and $P_{B}^{s}$}
\begin{figure}[hbtp]
\centering
\includegraphics[width=1\linewidth,height=0.25\textheight]{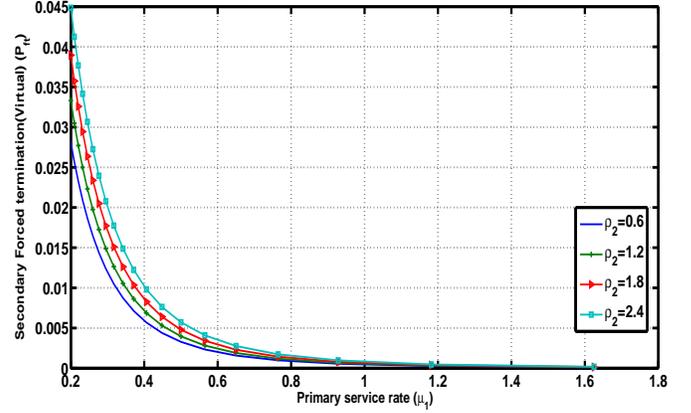}
\caption{$P_{FT}^{s}$ Vs. primary service rate $\mu_{1}$ at various secondary loadings}   \label{figmu1pft}
\label{fig.10}
\end{figure}

\begin{figure}[hbtp]
\centering
\includegraphics[width=1\linewidth,height=0.25\textheight]{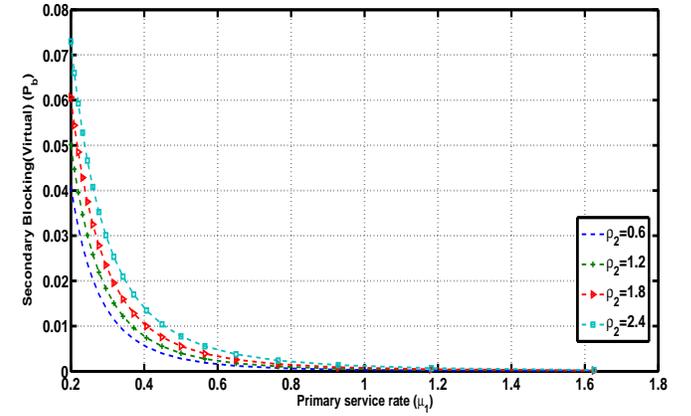}
\caption{$P_{B}^{s}$ Vs. primary service rate $\mu_{1}$ at various secondary loadings}   \label{figmu1pb}
\label{fig.11}
\end{figure}

In this section, we are concerned with the PU activity not its loading, i.e. how often do PUs depart with respect to a fixed arrival rate. Setting $\lambda_{p}=1.3$ and with the same previous system parameters, Figures \ref{figmu1pft} and \ref{figmu1pb} show the effect of changing $\mu_{1}$ on $P_{FT}$ and $P_{B}$, respectively. Obviously, as long as $\mu_{p}$ increases, the PU service time decreases, so he stays active for shorter times. Thus, the system rarely gets congested with PUs and consequently, the blocking and forced termination probabilities highly decrease. Hence, the activity of PUs on the network is an essential factor which effectively affects the SU behavior.

%\begin{figure}[hbtp]
%\centering
%\includegraphics[width=1\linewidth,height=0.25\textheight]{E:/variety/figures/thrvsm1virV2}
%\caption{Throughput Vs. primary service rate $\mu_{1}$ at various secondary loadings}   \label{figmu1thr}
%\label{fig.12}
%\end{figure}

\subsection{The effect of changing the minimum QoS requirements $C_{min}$ on $P_{FT}^{s} and P_{B}^{s}$}
\begin{figure}[hbtp]
\centering
\includegraphics[width=1\linewidth,height=0.25\textheight]{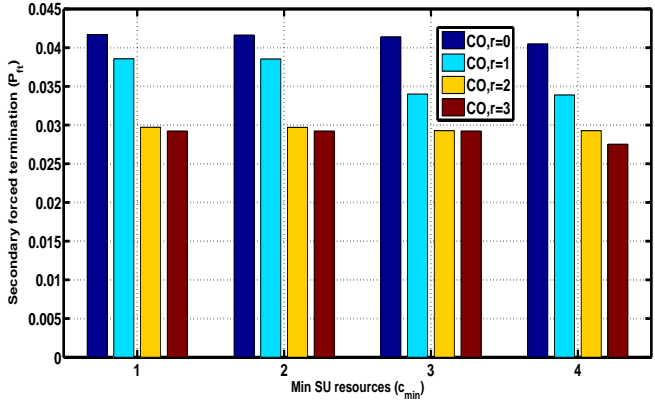}
\caption{$P_{FT}^{s}$ Vs. Min SU resources $C_{min}$ at various reservation cases}   \label{figcompare5}
\label{fig.13}
\end{figure}

\begin{figure}[hbtp]
\centering
\includegraphics[width=1\linewidth,height=0.25\textheight]{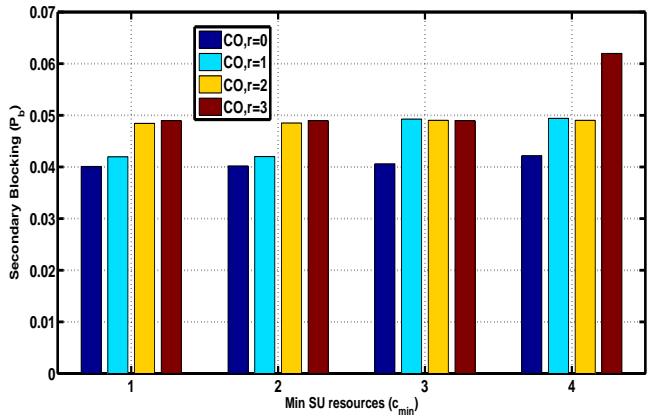}
\caption{$P_{B}^{s}$ Vs. Min SU resources $C_{min}$ at various reservation cases}   \label{figcompare6}
\label{fig.14}
\end{figure}

Definitely, changing $C_{min}$ leads to different network scenarios. In Figures \ref{figcompare5} and \ref{figcompare6}, we plot $P_{FT}^{s}$ and $P_{B}^{s}$, respectively, versus $C_{min}$ for different values of r. Apparently, increasing $C_{min}$ implies that fewer SUs would saturate the system. Hence, the SU blocking probability typically increases as $C_{min}$ increases. Additionally, the number of terminated SUs on the appearance of a PU would decrease, and so does the forced termination probability. On the reservation side, naturally, reserving more channels lowers the forced termination $P_{FT}^{s}$ but increases blocking new users. Such insights foster us to expect that systems with heavy real-time SU applications (larger bandwidth) are less susceptible to forcing SUs to terminate by PUs, however they result in blocking more new SUs and vice versa.

\section{Conclusion} \label{conclusion}
Spectrum mobility in cognitive networks is unavoidable to reduce the impact of SUs on PUs. In this paper, virtual reservation is introduced as a new link maintenance strategy to reduce the forced termination probability of SU sessions. Additionally, full spectrum utilization is proposed for 
greedy applications in collaborative cognitive networks. Our performance evaluation for virtual reservation and full spectrum utilization shows a noticeable improvement in key performance metrics including the SU throughput and secondary forced termination and blocking probabilities when compared to non-cooperative schemes or collaborative 
schemes that do not adopt full spectrum utilization. In addition, we study the system performance under many conditions such as changing $r$, $C_{min}$ and $\rho_{s}$, so that we could draw some insights about the entire cognitive model. Also, the novel approaches affirm that the SU throughput is clearly affected by the number of reserved channels, in contrast to what was concluded in preceding work \cite{pacheco2009optimal}. As a future work, we consider investigating dynamic reservation strategies instead of the presented static reservation strategies.
\bibliographystyle{IEEEbib}
\bibliography{IEEEabrv,myreference}
\end{document}